# Q-Drug: a Framework to bring Drug Design into Quantum Space using Deep Learning

Zhaoping Xiong [1*], Xiaopeng Cui [2*], Xinyuan Lin [1], Feixiao Ren [1], Bowen Liu [2], Yunting Li [3], Manhong Yung [2#] and Nan Qiao [1#]

[1] *Laboratory of Health Intelligence, Huawei Cloud Computing Technologies Co., Ltd, Guizhou, 550025, China*

[2] *Central Research Institute, Huawei Technologies, Shenzhen, 518129, China*

[3] *Institute for Nanoelectronic Devices and Quantum Computing, Fudan University, Shanghai 200433, China*

[*] Equal contribution

[#] Correspondence to: Nan Qiao (qiaonan3@huawei.com) and
Manhong Yung (yung.manhong@huawei.com)

## Abstract

Optimizing the properties of molecules (materials or drugs) for stronger toughness, lower toxicity, or better bioavailability has been a long-standing challenge. In this context, we propose a molecular optimization framework called Q-Drug (Quantum-inspired optimization algorithm for Drugs) that leverages quantum-inspired algorithms to optimize molecules on discrete binary domain variables. The framework begins by encoding the molecules into binary embeddings using a discrete VAE. The binary embeddings are then used to construct an Ising energy-like objective function, over which the state-of-the-art quantum-inspired optimization algorithm is adopted to find the optima. The binary embeddings corresponding to the optima are decoded to obtain the optimized molecules. We have tested the framework for optimizing drug molecule properties and have found that it outperforms other molecular optimization methods, finding molecules with better properties in 1/20th to 1/10th of the time previously required. The framework can also be deployed directly on various quantum computing equipment, such as laser pulses CIMs, FPGA Ising Machines, and quantum computers based on quantum annealing, among others. Our work demonstrates a new paradigm that leverages the advantages of quantum computing and AI to solve practically useful problems.

## Introduction

Molecular optimization is necessary when chemical molecules, such as drugs or materials, require improved properties such as lower toxicity, better bioavailability, and higher toughness. Modifying the structure of the molecules can achieve these outcomes, but it is a laborious and costly process with many trial-and-error attempts. Many machine learning models have been developed to optimize molecules, including conditional VAE[1,2], JT-VAE[3], and GCPN[4], which are representative models based on different optimization algorithms including conditional generation,

Bayesian optimization, and reinforcement learning, respectively. Although these methods have shown some progress, they still face many challenges, such as spending too much time and getting stuck in local minima, which negatively affect optimization outcomes. Additionally, previous molecular optimization methods focused on optimization over the continuous domain, which may not approach issues like activity cliffs of molecules appropriately. Because subtle changes of the structure of a molecule may lead to large differences in potency when activity cliffs occur[5]. We aim to extend molecular optimization methods to the discrete domain.

Quantum annealing is a discrete optimization process that uses quantum fluctuation characteristics and can find the global optimal solution when the objective function has many candidate solutions. The theory of quantum annealing dates back to 1989 when Ray et al. proposed that quantum tunneling could help escape from the local minima of classical Ising spin glasses with rugged energy landscapes[6]. The advantages of quantum annealing were verified in numerical tests in 1998[7], and in 1999, the first experiments on quantum annealing in LiHoYF Ising glass magnets were conducted[8]. D-Wave Systems marketed a quantum annealing machine built on superconducting circuits in 2011[9], and over the last decade, superconducting circuit-based and laser pulse-based quantum annealing machines have advanced rapidly with more qubits and better connections[10–16]. Quantum annealing begins with the quantum superposition of all possible states (candidate states) with the same weight, and then the physical system begins its quantum evolution based on the time-dependent Schrodinger equation. Quantum tunneling is generated between states according to the time-dependent strength of the transverse field, which causes the probability amplitudes of all candidate states to change continuously and achieve quantum parallelism[7]. Ultimately, the transverse field is closed, and the expected system obtains the solution to the original optimization problem that is, the corresponding ground state of the classical Ising model. D-Wave's quantum annealing system, a specialized quantum computing chip, can effectively solve Maxcut's combinatorial optimization (NP-hard) problems[9].

Besides quantum annealing machines, quantum-inspired optimization algorithms simulate quantum effects on classical computers and offer new types of quantum solutions. These algorithms leverage some quantum computing benefits on classical hardware to outperform traditional methods. They incorporate quantum mechanics principles, such as quantum fluctuation, quantum tunneling, and adiabatic quantum evolution, to avoid local optima[17–19]. A typical quantum-inspired optimization algorithm transforms the adiabatic quantum process into a classical dynamical process that preserves the adiabatic quantum evolution's features. By simulating this classical dynamical process, the algorithm can find the ground state configuration of the complex Hamiltonian, which corresponds to the global optimal solution of the complex objective function[20,21]. Simulated bifurcation (SB) is a numerical simulation of the adiabatic evolution of classical nonlinear Hamiltonian systems with bifurcation phenomena, where each nonlinear oscillator's bifurcation branches correspond to each Ising spin's two states[22]. SB is suitable for parallel computing because of its simultaneous updating and demonstrate 10 times faster than the laser-based coherent Ising machine (CIM) on an all-to-all connected 2000-node MAX-CUT problem[22].

To explore the possibility of quantum-based discrete optimization algorithms for molecular optimization, we can encode the molecules as binary embeddings, optimize over these embeddings, and then decode them back into molecules. Variational AutoEncoders (VAEs) are ideal candidates to achieve this goal. VAEs are a class of deep generative models that can encode data samples and generate (decode) samples based on new codes[23]. They consist of two terms, referred to as the reconstruction term and the regularization term. The reconstruction term ensures encoding accuracy, while the regularization term ensures that sampling from a distribution can generate (decode) valid data samples. The re-parametrization trick has made VAEs popular. This technique ensures sampling over a particular distribution and neural networks backpropagate correctly. Unlike VAEs that use normal distribution as the prior distribution, the VAEs in our research require a discrete Bernoulli distribution as the prior distribution. The continuous parametrization trick used by normal distribution is not applicable to our situation.

There are two main challenges. First, VAEs with discrete latent variables are difficult to train efficiently because backpropagation through discrete variables is generally not possible. A main approach to addressing this challenge is by using continuous relaxation. The basic idea is to find a distribution defined on the continuous domain that can approximate the required discrete distribution. Second, sampling from a factorial Bernoulli distribution for values that take either 0 or 1 with equal 0.5 probability is too noisy to reconstruct valid data samples. Thus, we introduce a Restricted Boltzmann machine (RBM) as a generative model that learns the distribution of the encoder output and reduces the noise in the sampling codes.

With the improvements made through the use of continuous relaxation and Restricted Boltzmann Machines (RBM), molecules can now be effectively encoded as vectors in binary form (binary embeddings). And sampling from a Bernoulli distribution enables valid molecules to be decoded. By gathering a batch of molecules' properties, we can construct a predictive matrix factorization model for this property with the binary embedding of these molecules. Matrix factorization naturally takes the form of an Ising problem which can be solved efficiently by quantum annealing. However, in this study we used a quantum-inspired optimization algorithm called Simulated Bifurcation (SB) to accelerate this combinatorial optimization problem. The origin of SB is the Ising problem, which involves finding a spin configuration that minimizes energy (Ising Hamiltonian). Solving the Ising problem with the SB algorithm results in the optimal value and the optimal spin configuration (binary embedding). Decoding the binary embedding results in the obtaining optimal molecules with the corresponding property. In summary, we have constructed a framework that encodes molecules into binary embedding, which can be used to construct an Ising-like function to fit the properties of the molecules. The function can be further used as an objective function of the property optimization problem. By using quantum annealing with quantum computers or quantum-inspired algorithms on traditional computers, we can find better global optima and use less time compared to using traditional molecular optimization methods.

Overall, as illustrated in Figure 1, our methodology involved pretraining the Discrete VAE on a vast dataset of druglike molecules called ZINC-250k[24]. This step facilitated the encoding of molecules into binary codes that decoded into valid and druglike molecules. Using a quantum-inspired optimization algorithm, we manipulated the binary codes to minimize or maximize objective functions, thereby optimizing the molecules in the binary domain.

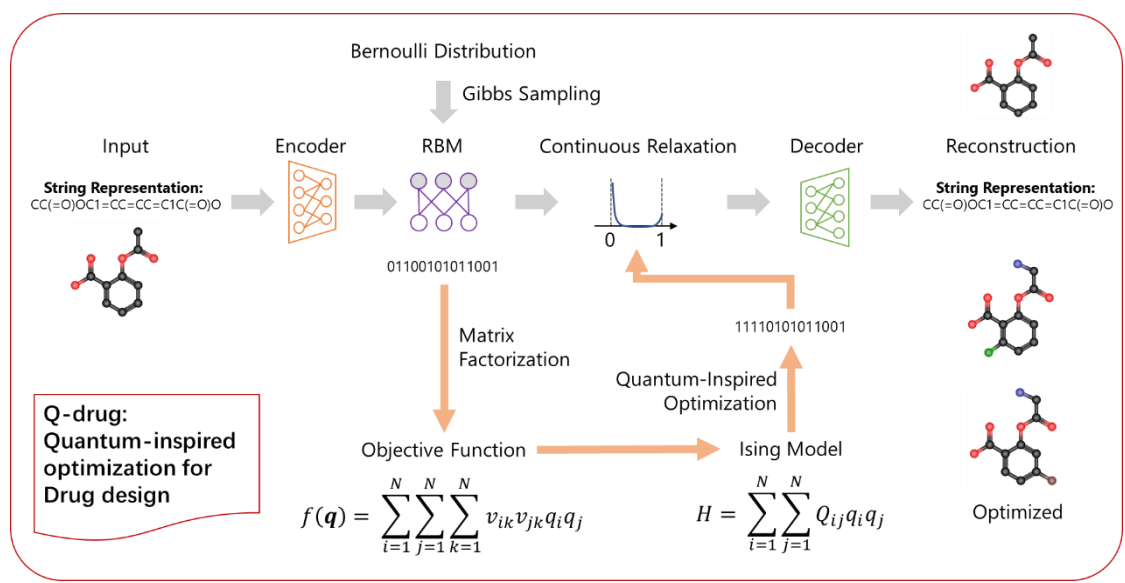

**Figure 1. The framework of Quantum-Inspired Molecular Optimization (Q-drug).** The path indicated by the gray arrows is the pre-training path that ensures the binary codes decode to druglike and valid molecules. The path indicated by orange arrows is the optimization path that optimize the molecules in binary domain with quantum-inspired optimization algorithm.

## Methods

### Binary encoding with variational autoencoders

Variational AutoEncoders (VAEs) are a class of deep generative models that can encode data sample into latent space embeddings and generate(decode) samples based on random sampled new embeddings[23]. It consists of an encoder and a decoder. The encoder of our experiments is composed of a three-layer gated recurrent units (GRUs)[25], which takes the Self-Referencing Embedded Strings (SELFIES) representation of molecules as input[26]. The decoder is composed of a one-layer GRU, which outputs SELEFES string representation of molecules. The latent variable of VAE is the representation of the molecule. We require the representation to be binary embeddings and the model to train efficiently.

### Continuous relaxation

Following the discrete VAE[27], we use the spike-and-exponential function to transform the discrete binary z to continuous $\zeta$:

$$r(\zeta|z=0) = \begin{cases} \infty, & \zeta = 0 \\ 0, & \zeta \neq 0 \end{cases} \qquad F_{r(\zeta|z=0)}(\zeta') = 1, \qquad (1)$$

$$r(\zeta|z=1) = \begin{cases} \dfrac{\beta e^{\beta\zeta}}{e^{\beta}-1}, & 0 < \zeta \leq 1 \\ 0, & \zeta \leq 0 \text{ or } \zeta > 1 \end{cases} \qquad F_{r(\zeta|z=1)}(\zeta') = \dfrac{e^{\beta\zeta}}{e^{\beta}-1}\bigg|_0^{\zeta'} = \dfrac{e^{\beta\zeta'}-1}{e^{\beta}-1}, \qquad (2)$$

Where $r(\zeta)$ is the probability function, and $F_{r(\zeta)}(\zeta')$ is the corresponding CDF (cumulative distribution function). When z=0, ζ has a Dirac delta distribution, which is zero everywhere except at zero (Equation 1). The value at zero is positive infinite to ensure the integral over the entire real line is equal to one. When z=1, ζ has an exponential distribution in the domain (0, 1]. This transformation from z to ζ is invertible: ζ = 0 ⇔ z = 0, and ζ > 0 ⇔ z= 1 almost surely. As shown in Figure 1 inset of continuous relaxation, this function is a good approximation.

## Re-parametrization on the relaxed distribution

To sample effectively from the relaxed continuous distribution, a common method is to sample from the inverse CDF whose domain is [0, 1], using a uniform random variable $u \sim U[0,1]$ (Equation 3). However, the binary condition z of CDF is unknown, so an RBM is introduced to generate the binary condition z. We require the RBM to generate binary z that approximates the latent variable q of VAE. Here, RBM evolves via block Gibbs sampling and gradient descent.

To sample based on the latent variable $q$ of VAE, we need to re-parametrize q with $\rho \sim U[0,1]$ as done in Equation 4. The embedding is binarized to 1 when $q \geq 1 - \rho$, and to 0 when $q < 1 - \rho$.

$$\zeta' = \begin{cases} \frac{1}{\beta} \cdot \log[u \cdot (e^\beta - 1) + 1], & z = 1 \\ 0, & z = 0 \end{cases} \tag{3}$$

$$\zeta' = \begin{cases} \frac{1}{\beta} \cdot \log\left[\left(\frac{q + \rho - 1}{q}\right) \cdot (e^\beta - 1) + 1\right], & q \geq 1 - \rho \\ 0, & q < 1 - \rho \end{cases} \tag{4}$$

## Quantum-inspired optimization

Simulated bifurcation is a heuristic algorithm inspired by quantum physics theory that aims to find an optimal spin configuration that minimizes the Ising energy[17,20]. It is a classical analogue of quantum adiabatic bifurcation in nonlinear oscillators.

$$E_{Ising} = \sum_{i}^{N} \sum_{j}^{N} J_{ij} s_i s_j, \tag{5}$$

Where N is the number of spins, $s_i$ denotes the ith spin of value +1 or −1, and $J_{ij}$ is the coupling coefficient between the ith and jth spins.

Ballistic SB (bSB) represents the state-of-the-art quantum-inspired optimization algorithm. The equation of motion of bSB read:

$$\begin{aligned} \dot{x}_i &= \frac{\partial H_{bSB}}{\partial y_i} = a_0 y_i, \\ \dot{y}_i &= \frac{\partial H_{bSB}}{\partial x_i} = [a_0 - a(t)] x_i + c_0 \sum_{j}^{N} J_{ij} x_j, \end{aligned} \tag{6}$$

Where $x_i, y_i$ are the positions and momentums of a particle corresponding to the ith spin, $a(t)$ is a control parameter increased from zero to $a_0$, and $a_0, c_0$ are positive constants. The Hamiltonian HbSB reads:

$$H_{bSB} = \frac{a_0}{2} \sum_{i}^{N} y_i^2 + V_{bSB}, \tag{7}$$

$$V_{bSB} = \begin{cases} \dfrac{a_0 - a(t)}{2} \sum_{i}^{N} x_i^2 - \dfrac{c_0}{2} \sum_{i}^{N} \sum_{j}^{N} J_{ij} x_i y_j, & |x_i| \leq 1 \\ \inf, & |x_i| > 1 \end{cases} \tag{8}$$

Where the kinetic energy related to momentum y and potential energy $V_{bSB}$ (Equation 8) is summed. Inelastic walls are set at $x_i = \pm 1$, that is, if $|x_i| > 1$ during the evolution, replace $x_i$ with $sgn(x_i)$ and set $y_i = 0$.

## Objective function construction

In order to optimize the specific attributes of molecules, we need some molecules with known properties to construct the objective function $f(\boldsymbol{q})$ of the optimization model. In this solution, a matrix factorization method is used to construct a predictive function of the property. The function form is as follows:

$$f(\boldsymbol{q}) = \sum_{i=1}^{N} \sum_{j=1}^{N} \sum_{k=1}^{N} v_{ik} v_{jk} q_i q_j \tag{9}$$

Equation 9 shows the form of objective function $f(\boldsymbol{q})$, where $q_i$ and $q_j$ are values of the ith and the jth dimension of the binary embeddings of a molecule, and $v_{ik}$ and $v_{jk}$ are coefficients of the kth factor. The objective function $f(\boldsymbol{q})$ is constructed to predict the property value of the molecule using a matrix factorization model. We first input the binary embeddings of molecules and determine the coefficients ($v_{ik}$ and $v_{jk}$) by fitting the properties of molecules with matrix factorization. With fixed $v_{ik}$ and $v_{jk}$, the function $f(\boldsymbol{q})$ has the same form as the Hamiltonian of the Ising model (as follows):

$$H_{problem} = \sum_{i=1}^{N} \sum_{j=1}^{N} Q_{ij} q_i q_j \tag{10}$$

Equation 10 defines the Ising model Hamiltonian, where $q_i$ and $q_j$ are the spin states of the ith and jth elements, and $Q_{ij}$ is the coupling coefficient between them. The Ising energy-like function is obtained by summing up the Equation 9's $v_{ik}$ and $v_{jk}$ along dimension k to get $Q_{ij}$. With fixed $Q_{ij}$, the objective function $f(\boldsymbol{q})$ (or the Ising model) can get the ground state by using quantum annealing. The corresponding energy of the ground state is the predicted property value and decoding the ground state (which is also the binary embeddings) will yield the molecules with optimal properties.

# Results

## Discrete VAE can effectively encode and decode molecules

We want to design a good encoding scheme for molecules, which can transform molecules into binary embeddings and reconstruct valid molecules from random binary embeddings. To establish a baseline, we experimented with the widely-used VAE (Variational Autoencoder) to encode molecules into continuous embeddings. The dataset we use is ZINC-250k, which includes about 250, 000 druglike molecules. A simple and intuitive extension of VAE is Bernoulli VAE, which directly changes the gaussian distribution of prior latent space to Bernoulli distribution. Discrete VAE is a more complex model that uses continuous relaxation and RBM (Restricted Boltzmann Machine) generation.

Table1. Comparison of different encoding schemes

| Model | VAE (z =128) | Bernoulli VAE (z = 128) | Bernoulli VAE (z = 2048) | Discrete VAE (z = 128) | Discrete VAE (z = 2048) |
|---|---|---|---|---|---|
| Reconst. Loss (↓) | **0.007** | 0.508 | 0.310 | 0.202 | 0.008 |
| KL Divergence (↓) | 0.001 | 0.000 | 0.000 | 0.172 | **-2.969** |
| Ratio of Recon. (↑) | **94.2%** | 0.00% | 0.10% | 12.4% | 91.7% |
| Validity (↑) | 100% | 100% | 100% | 100% | **100%** |
| Uniqueness (↑) | 95.2% | 80.7% | 92.3% | 99.5% | **99.9%** |
| Novelty (↑) | 93.8% | 70.9% | 74.8% | 89.6% | **99.9%** |

The best values of the metrics across different encoding schemes are in bold.

Table 1 shows that decoding validity is easy to achieve because our model uses the SELFIES string of molecules as input, which guarantees that almost every string corresponds to valid molecules. The reconstruction ratio reveals that Bernoulli VAE can hardly reconstruct the same molecule from its binary embedding, even if we increase the latent dimension z from 128 to 2048. This means that the binary embedding does not capture the molecular structure well, which indicates that Bernoulli VAE is not suitable for our situation. On the other hand, discrete VAE performs comparably to continuous VAE on almost all metrics and slightly outperforms it when we enlarge the latent z.

## Q-drug outperform previous molecular optimization models on different tasks.

Table 2. Comparison of different molecular optimization models

| Tasks | CVAE Conditional | JT-VAE Bayesian | GCPN RL | Q-drug Quantum |
|---|---|---|---|---|
| QED (↑) | 0.882 | 0.904 | **0.948** | **0.948** |
| FGFR3 [pIC50] | 9.502 | 9.817 | 10.555 | **10.690** |
| +10* QED (↑) | +10*0.781 | +10*0.840 | +10*0.631 | **+10*0.840** |
| FGFR4 [pIC50] | 9.572 | 10.157 | 10.851 | **10.813** |
| +10* QED (↑) | +10*0.803 | +10*0.760 | +10*0.684 | **+10*0.904** |
| Ave. infer. time | ~5s | ~15min | ~30min | ~90s |

The best values of the metrics across different encoding schemes are in bold. We report the largest value each model can achieve.

We compare three optimization models for molecular optimization: conditional VAE, JT-VAE and GCPN, which use conditional generation, Bayesian optimization and reinforcement learning respectively. On the QED drug-likeness optimization task, Q-drug and GCPN can achieve a QED value of 0.948, which is the practical upper bound of QED value although it ranges from 0 to 1. However, if we examine the molecules in Figure 2C, we find that GCPN may just exploit the flaw of QED to get high QED, because the molecule is obviously not ideal for researchers with medicinal chemistry background. In contrast, CVAE, JT-VAE and Q-drug generate more reasonable structures, and Q-drug has the best optimization effect. When comparing the inference time cost of these models, it is observed that the conditional VAE is fast, but its optimization effect is inferior. On the other hand, Q-drug takes only about 1/20th to 1/10th of the time required for Bayesian optimization and reinforcement learning to achieve the best optimization effect.

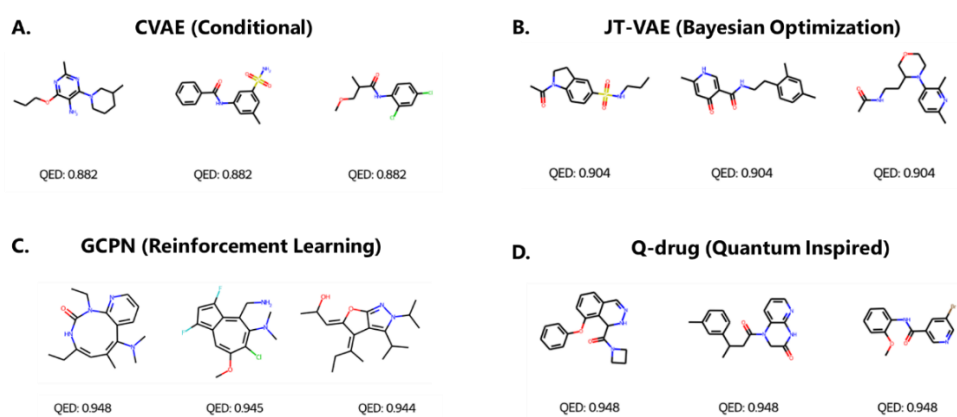

Figure 2. Samples of generated molecules in Drug-likeness (QED) optimization with four different methods.

For multi-objective optimization tasks that optimize bioactivity and QED drug-likeness simultaneously, we scale both metrics to the same range by adding 10 times QED value to pIC50 value. We observe that Q-drug with quantum-inspired optimization method outperforms the other methods on both drug targets: FGFR3 and FGFR4 tasks (Table 2, Figure 3 and Figure 4).

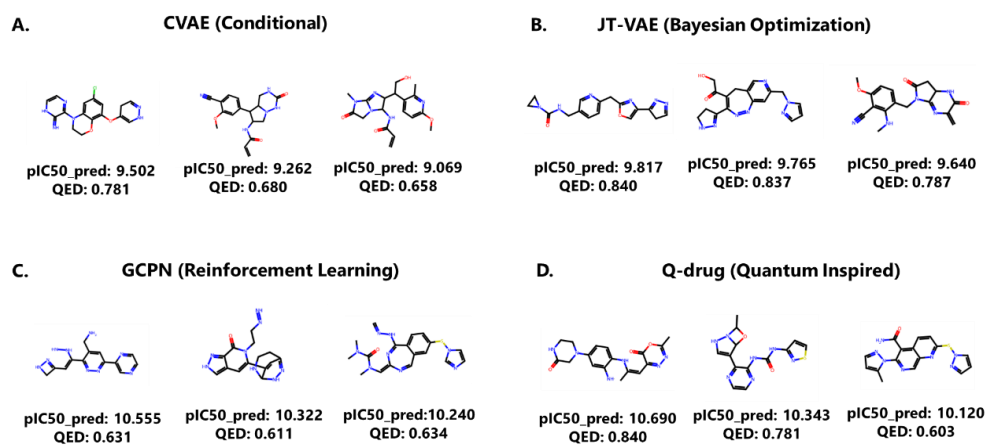

Figure 3. Samples of generated molecules in FGFR3 bioactivity and QED optimization with four different methods.

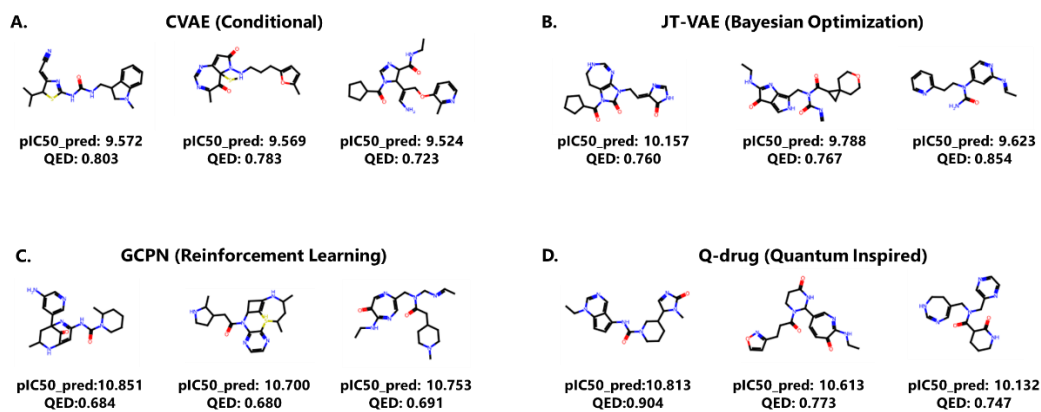

Figure 4. Samples of generated molecules in FGFR4 bioactivity and QED optimization with four different models.

## Discussion

The world around us is quantized or discrete, meaning that it has distinct, indivisible units. However, conventional molecular optimization methods have focused on optimizing over the continuous domain, which might not appropriately address issues such as "activity cliffs" in molecules. Our work explores the feasibility and promise of optimizing molecules over the discrete domain. We demonstrate a proof of concept using a discrete VAE to encode molecules, but also acknowledge the potential for further improvement using techniques such as DVAE++[28]. To solve the constructed Ising energy-like objective function, we propose utilizing quantum computing-based equipment in addition to quantum-inspired optimization algorithms. This includes superconducting quantum computers[9,11], laser pulses based CIM (coherent Ising machines)[10,12–16], and FPGAs (field-programmable gate arrays) based Ising machines[29–31]. Constructing objective functions with HOFM (Higher-Order Factorization Machines)[32] or BOX-QUBO (Black box Optimization using the Cross (X) entropy method and QUBO)[33] also shows potential and is suitable for optimizing by quantum-inspired algorithms. Our research provides a practical framework called Q-drug, which brings drug design into the quantum space utilizing AI, offering novel possibilities for better molecular design techniques based on quantum computing concepts. Furthermore, other applications such as molecular conformation generation, and molecular docking may also significantly benefit from optimization through quantum-inspired algorithms.